\documentstyle[prl,aps,epsfig,multicol]{revtex}

\begin{document}
\title{Intercalant-Driven Superconductivity in YbC$_{6}$ and CaC$_{6}$}
\author{I.I. Mazin}
\address{Center for Computational Materials Science,\\
Naval Research Laboratory, Washington, DC 20375}
\date{\today }
\maketitle

\begin{abstract}
Recently deiscovered superconductivity in YbC$_6$ and CaC$_6$ at temperatures 
substantially higher than previously known for intercalated graphites, 
raised several new questions: (1) Is the mechanism considerably different
from the previously known intercalated graphites? (2) If superconductivity
is conventional, what are the relevant phonons? (3) Given extreme similarity
between YbC$_6$ and CaCa$_6$, why their critical temperatures are so different?
We address these questions on the basis of first-principles calculations  and
conclude that
 coupling with intercalant phonons is likely to be the main force for 
superconductivity  in YbC$_6$ and CaC$_6$, but not in alkaline-intercalated
compounds, and explain the difference in $T_c$ by the ``isotope effect'' due
to the difference in Yb and Ca atomic masses.
\end{abstract}


\begin{multicols}{2}

Recent discovery of relatively high temperature superconductivity in
graphite intercalated compounds (GIC) YbC$_{6}$ and CaC$_{6}$\cite{exp} of
6.5 and 11.5 K, respectively, the highest among GIC, has renewed theoretical
interest in superconductivity in GIC\cite{teor,we}. In particular, it
inspired Csanyi $et$ $al$\cite{teor} to analyze four superconducting and
three non-superconducting GIC in order to elucidate common trends and get
more insight into the mechanism of superconductivity. They discovered an
interesting
empirical correlation between the occupation of the only 3D band in the
system, and the appearence of superconductivity, and, using this
observation, they argued that superconductivity in all GIC is
electronic by origin, intermediate bosons being probably excitons or acoustic
plasmons. This calls for revising the conventional wisdom that
superconductivity in GIC is conventional by nature and mostly due to carbon
phonons.

In this Letter we shall argue, using first principle calculations
and experimental data \cite{exp,belash}, that while the standart picture of
electron-phonon coupling mainly with the C modes is probably in doubt, at
least in these two compounds, superconductivity is likely to arise from the
inetercalant vibrations, and not from electronic excitations. In this sense,
YbC$_{6}$ and CaC$_{6}$ are somewhat close to another high-T$_{c}$ (18K)
transition metal - carbon superconductor, Y$_{2}$C$_{3},$ where
superconductivity seems to be related to Y phonons\cite{y2c3}

Our analysis is based on highly accurate all electron fully relativistic
LAPW calculations \cite{WIEN}. LDA+U correction was applied to the
f-electrons in Yb, to account for Hubbard correlations. Details of the
calculations for YbC$_{6}$ are described elsewhere\cite{we}. Calculations
for CaC$_{6}$ and for other materials discussed below
were performed in the same setup as for YbC$_{6},$ but without
LDA+U and spin-orbit corrections. For the purpose of comparison, we also
performed similar calculations for Li GIC: LiC$_{6}$ and LiC$_{3}.$

Let us first discuss the viability of the electronic mechanism scenario\cite%
{teor}. This conjecture is based on four assumptions: (1) the 3D free
electron like band crosses the Fermi level in all superconducting GIC and
is fully empty in all nonsuperconducting ones; (2) this band is not related
to intercalant $s$ or $p$ states, but is formed by free electrons propagating in
the interstitial space; (3) this band is much weaker coupled with the phonons than
the other bands, and (4) such band stracture is advantageous for the excitonic
"sandwich" mechanism\cite{sand} or for the acoustic plasmons mechanism\cite{ap}. 

The first
assumption is correct for many, but, apparently, not all GIC. For instance, in LiC$_3$, in
pseudotential calculations of Ref. \cite{teor}, the band in question touches
the Fermi level. In our fully converged all-electron calculations with a
fine k-mesh (13x13x10) this band was 0.2 eV above the Fermi level (Fig. \ref%
{bandsLi}). Yet, according to the experiment, superconductivity was observed
in this compound\cite{belash}, although because of low temperature and broad
transition the authors failed to give the exact number for $T_{c}.$ On the
other hand, Eu in EuC$_{6}$ is known to be divalent\cite{SM}, just as Yb or Ca,
and forms exactly the same crystal structure, yet the
material is not superconducting \cite{MarkE}. Eu in EuC$_{6}$ is magnetic,
 but, if Eu electrons are not involved in superconductivity,
 the long coherence length in GIC would have prevented magnetic
pair-breaking, as long the material remains well ordered
antiferromagnetically ($cf.$, for instance, superconducting antiferromagnetic 
Chevrel phases).
\begin{figure}[tbp]
\centerline{\epsfig{file=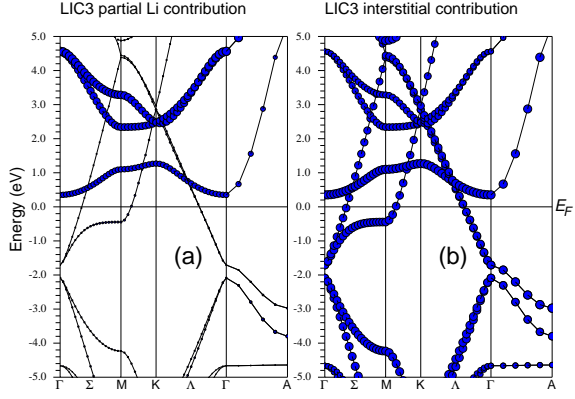,width=0.95\linewidth,angle=0,clip=}}
\vspace{0.3cm}
\caption{LAPW band structure of LiC$_3$.
The left panel shows the partial Li character, and the right panel interstitial
character. Note uniform participation of the interstitial states in all bands,
and selective participation of the Li states in the free electron like band.
(color online) }
\label{bandsLi}
\end{figure}

\begin{figure}[tbp]
\centerline{\epsfig{file=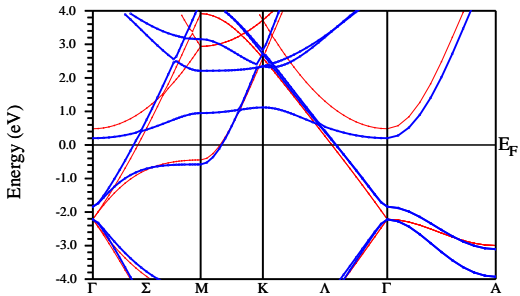,width=0.95\linewidth,angle=0,clip=}}
\vspace{0.3cm}
\caption{Band structure of LiC$_3$ (thick blue lines) and $e$C$_3$
(thin red lines). 
Note that the most affected band is the free electron like one, specifically,
its in-plane dispersion.
(color online) }
\label{empty}
\end{figure}

 The second assumption is somewhat philosophical,
because it is hard to tag an itinerant free-electron like
band as an interstitial or as an $sp$ band of
an alkali metal. However, decomposition of the wave function of this band
shows (for instance,
in case of LiC$_{3}$, displayed in Fig. \ref{bandsLi}), that while interstitial
plane wave states have the same weight in this band as in the other, 2D
states, Li $s$ and $p$ (mostly $p_{z})$ orbitals participate nearly
exclusively in this band, and provide much more share of the total weight
than the volume occupied by the Li MT spheres. By the standard band theory
parlance, this identify them as at least substantially Li-derived. 
As an independent test, we
performed calculations for a hypothetical compound in which the Li atom
were replaced by a free electron, and found that the 3D band dispersion
changed enormously (Fig.\ref{empty}).

\begin{figure}[tbp]
\centerline{\epsfig{file=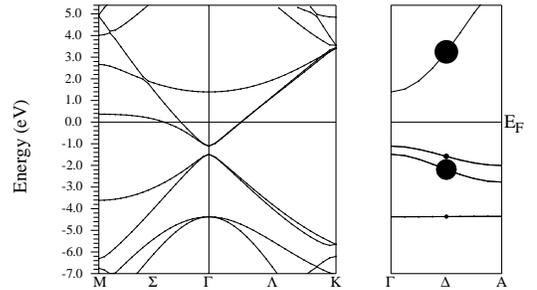,width=0.95\linewidth,angle=0,clip=}}
\vspace{0.3cm}
\caption{LAPW band structure of LiC$_3$. The radii of the filled 
circles half-way between $\Gamma$ and A are proportional to the 
electron-phonon interaction matrix elements of the correcponding 
band with the Li breathing mode.
Note that the most affected band is the free electron like one.
}
\label{phonsLi}
\end{figure}

The validity of the third assumption can be tested by direct calculations:
one can evaluate the electron-phonon matrix elements at a particular 
high-symmetry point in the Brillouin zone with a specific phonon by applying 
a frozen displacement and looking at the induced band splittings. We 
employed this technique to compute the coupling at the point half-way between 
$\Gamma$ and A with the ``breathing'' Li phonon, that is, the one 
corresponding to 
a breathing displacement of Li along $c$. The results for LiC$_6$
are  shown in Fig. \ref{phonsLi}. One can see that within $\pm$ 5 eV of the Fermi 
level the free-electron like band in the one with the  {\it 
strongest } electron-phonon coupling.
In view of the noticed in Ref. \cite{teor} high
sensitivity of this band to interplane distance, one should also expect
a strong coupling with the buckling C modes, but we did not test this numerically.

 Turning to the "sandwich" mechanism\cite{sand}, we
observe that the original papers were strongly based on the idea that the
electonic excitations reside in a {\it dielectric }layer (otherwise metallic
screening prevents exciton formation), while the interlayer band is 
{\it metallic} in the well superconducting GIC, as observed in Ref. \cite{teor}.
 Finally, acoustic plasmons
would form in this band either if its effective mass were much heavier than
in the other bands, or if it were 2D. Neither condition holds.
 
\begin{figure}[tbp]
\centerline{\epsfig{file=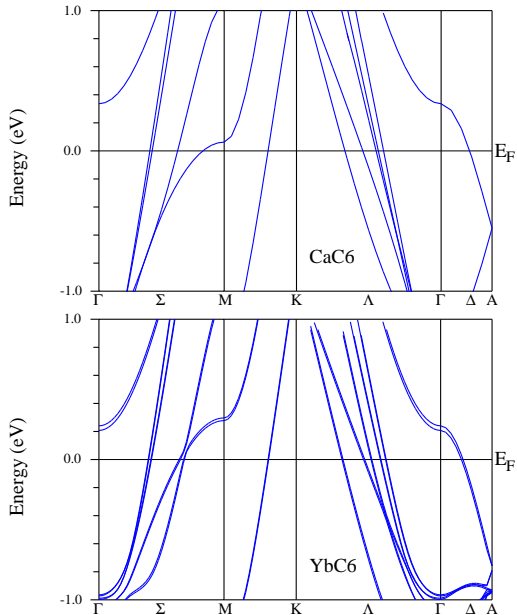,width=0.95\linewidth,angle=0,clip=}}
\vspace{0.3cm}
\caption{Band structure of CaC$_6$ near the Fermi level,
compared with that of YbC$_6$.  }
\label{Ca-Yb}
\end{figure}

\begin{figure}[tbp]
\centerline{\epsfig{file=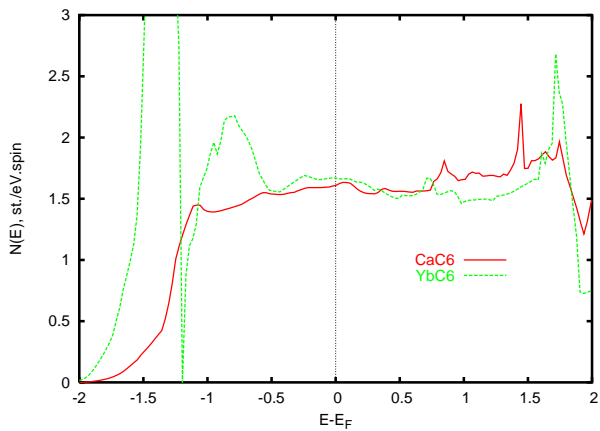,width=0.95\linewidth,angle=0,clip=}}
\vspace{0.3cm}
\caption{Density of states  of CaC$_6$ near the Fermi level,
compared with that of YbC$_6$. (color online) }
\label{DOS}
\end{figure}

Since electronic superconductivity appears to be rather unlikely, we need to find
another mechanism. It was conjectured in Ref. \cite{we} that
superconductivity in YbC$_{6}$ is largely due to Yb phonons, in analogy with
Y$_{2}$C$_{3}$\cite{y2c3}. Comparison between YbC$_{6}$ and CaC$_{6}$ lends
additional support to this scenario. Indeed, a detail examination of the two
band structures (Fig. \ref{Ca-Yb})
finds
 practically no difference  for all but one band in the
vicinity of the Fermi level, including the 3D
interlayer band. There is some effect of additional hybridization 
with the f-states on the lowest unoccupied state at the M point,
which, however, affects one band out of six, and does not change
the  density of states near the Fermi level
(Fig.\ref{DOS}).
 If superconductivity were not related to the intercalant
atom, one would expect the critical temperature to change only slightly,
 a situation 
analogous to YBa$_{2}$Cu$_{3}$O$_{7},$ where Y can be substituted by any
trivalent rare-earth with $T_{c}$ changing within a few per cent only. On
the contrary, critical temperature of CaC$_{6}$ is 1.77 of that of YbC$_{6}. 
$

At this point we observe that $\sqrt{M_{Yb}/M_{Ca}},$ where $M_{Ca(Yb)}$ is
the atomic mass of Ca(Yb), is 2.08. This means that ``isotope effect'' on $%
T_{c},$ due to substitution of Yb by Ca, is 1.77/2.08=0.85 of the ``full''
isotope effect if superconductivity were entirely due to Yb/Ca modes, and 
no other differences between the two materials was relevant for superconductivity.
Recalling that partial isotope effects in binaries are scaled with partial
coupling constants, we find that $\lambda _{R}/(\lambda _{R}+\lambda
_{C})\approx 0.85,$ (here $R$ stands for either Ca or Yb), that is, 15\% of
the electron-phonon coupling comes from C, and the rest from Ca/Yb. It is
more curious than important that the rough estimate given in Ref. \cite{y2c3}
for Y$_{2}$C$_{3}$ was 10\% of total coupling coming from C-C phonons, and the 
rest from 
 pure Y or mixed Y-C modes, in an interesting agreement
with the above estimate for YbC$_{6}.$

To summarize, we propose that unusually high for intercalated graphites
critical temperatures in CaC$_{6}$ and YbC$_{6}$ are mainly due to
substantial participation of the intercalant electronic states at the Fermi
level, and, as a consequence, sizeable coupling with soft intercalant modes.
It remains unclear to what extent the same mechanism is present in other, low $%
T_{c}$ GIC, such as KC$_{x},$ LiC$_{x}$ and NaC$_{x}.$ Although their
electronic structure shares some similarities with YbC$_{6}/$CaC$_{6},$ it
is substantially different, especially regarding intercalant states. It
seems unlikely that intercalants in the former are involved in
superconductivity nearly as strong as in the latter.

Finally, let us discuss what experiments can test the proposed scenario.
Measuring isotope effect on Ca is predicted to yield an exponent of the
order 0.4, and that on C of 0.1 or less. Another prediction is that mixed
intercalation of Ca and Yb should produce samples whose $T_{c}$ scales with
concentration as the average logarithmic phonon frequency, that is, as $%
T_{Ca}^{x}T_{Yb}^{(1-x)},$ where $x$ is the Ca concentration. An interesting
question is, what would be a result of partial substitution of Ca with Mg or
Sr? Their ionic radii are substantially different from those of Ca or Yb
(which are practically the same in hexagonal coordination). A moderate
substitution with, say, Mg will reduce the interplanar distance, thus making
Ca-C force constants larger and the coupling constant with electrons for Ca
modes smaller. On the other hand, Mg ions themself will sit a pore relatively
large for their ionic radius, and thus will have smaller force constant,
leading to some increase of $\lambda .$ The third effect is that the
corresponding Mg modes will have higher frequency for the same force
constants because of smaller mass. If the first two effect approximally
cancel each other, co-doping with Mg may be a route to even higher $T_{c}.$
Obviously, more experimental and computational work is required to clarify
this issue.

This was supported by the Office of Naval Research.

\end{multicols}

\end{document}